\documentclass[
preprint,
]{revtex4-1}

\usepackage{amsthm}
\usepackage{color}
\theoremstyle{plain}
\newtheorem{theorem}{Theorem}
\newtheorem*{theorem*}{Theorem}

\newtheorem*{definition*}{Definition}

\newtheorem*{lemma*}{Lemma}

\newtheorem{remark}[theorem]{Remark}

\usepackage{txfonts}

\newcommand{\be}{\begin{eqnarray}}
\newcommand{\ee}{\end{eqnarray}}
\newcommand{\no}{\nonumber}

\newcommand{\Tr}{\mbox{Tr} }

\begin{document}
\title{Free energy equivalence between mean-field models and nonsparsely diluted mean-field models }
\author{Manaka Okuyama$^1$\footnote{corresponding author: manaka.okuyama.d2@tohoku.ac.jp}}
\author{Masayuki Ohzeki$^{1,2,3}$}
\affiliation{$^1$Graduate School of Information Sciences, Tohoku University, Sendai 980-8579, Japan}
\affiliation{$^2$Department of Physics, Institute of Science Tokyo, Tokyo 152-8551, Japan}
\affiliation{$^3$Sigma-i Co., Ltd., Tokyo 108-0075, Japan} 

\begin{abstract} 
We studied nonsparsely diluted mean-field models that differ from sparsely diluted mean-field models, such as the Viana--Bray model.
When the existence probability of each edge follows a Bernoulli distribution, we rigorously prove that the free energy of nonsparsely diluted mean-field models with appropriate parameterization coincides exactly with that of the corresponding mean-field models in ferromagnetic and spin-glass models composed of any discrete spin $S$ in the thermodynamic limit.
Our results is a broad generalization of the result of a previous study [Bovier and Gayrard, J. Stat. Phys. 72, 643 (1993)], where the densely diluted mean-field ferromagnetic Ising model (diluted Curie--Weiss model) with appropriate parameterization was analyzed rigorously, and it was proven that its free energy was exactly equivalent to that of the corresponding mean-field model (Curie--Weiss model).

\end{abstract}
\date{\today}
\maketitle

\section{Introduction}
The importance of understanding the effects of randomness on the system properties has increased over the past few decades.
Randomness is widely employed to model and analyze phenomena in statistical mechanics and information science, such as in random matrix theory~\cite{TV}, stochastic differential equation~\cite{Sobczyk}, and random network~\cite{FK}.
The spin-glass model is a representative example of randomness in statistical mechanics, characterized by random interactions between spins.
The mean-field theory of spin glass models exhibits a nontrivial and rich structure, even though it is the starting point for analyzing random spin systems.
The concept of replica symmetry breaking~\cite{Parisi} in mean-field spin glass models has had a remarkable impact on both statistical mechanics and information science~\cite{Nishimori,MM}.
Mean-field spin glass models continue to be the focus of active research~\cite{CMMPTSZ}.

In statistical mechanics, randomness can be introduced not only by randomizing the interaction strength, as in spin-glass models, but also by diluting the interaction.
For the system size $N$, diluted mean-field models can be defined in three ways:
The first definition is sparsely diluted mean-field models, where the strength of the interaction is $O(1)$ and the existence probability of each edge is $O(N^{-1})$.
A concrete example of sparsely diluted mean-field models is the spin model on the Erd\H{o}s-R\'enyi random graph (such as the Viana--Bray model~\cite{VB}).
The sparsely diluted mean-field models are closely related to information science problems ~\cite{MZKST,FLTZ,MPZ,MM}.

The second definition of diluted mean-field models considers a densely diluted mean-field model, where the strength of the interaction is $O(N^{-1})$ and the existence probability of each edge is $O(1)$. 
Densely diluted mean-field models are naturally defined from a statistical mechanics perspective. As the number of interactions is proportional to $O(N^{2})$, the properties of these models are expected to be similar to those of the corresponding mean-field model.
Bovier and Gayrard~\cite{BG} proved that the free energy of the densely diluted Curie--Weiss model with appropriate parameterization coincides exactly with that of a Curie--Weiss model in the thermodynamic limit.
Densely diluted mean-field models have not been studied well~\cite{Sompolinsky,CN,MT}, but progress has been made recently~\cite{BCFBR,KLS,KLS2,KLS3,BMP,Boettcher,Wang}.
The zero-temperature properties of the densely diluted Sherrington--Kirkpatrick (SK) model (not the Viana--Bray model) were numerically investigated in Refs. \cite{Boettcher}. It was revealed that the ground-state energy coincides with that of the SK model and depends neither on the distribution of interactions nor on the concentration of dilution.
Interestingly, this universal behavior appears to be within the limit of the dilution concentration $\alpha\to0$.
However, the thermodynamic properties have not been clarified at finite temperatures.

The third definition of diluted mean-field models considers an intermediate regime between sparse and dense dilution, where the strength of the interaction is $O(N^{-b})$ and the existence probability of each edge is $O(N^{b-1})$ with $0<b<1$. 
Recent studies~\cite{MP,FM} showed that the thermodynamic properties of this intermediate regime are equivalent to those of the corresponding mean-field models if the existence probability of each edge follows a Bernoulli distribution.
In addition, if the existence probability of each edge follows an exponential distribution, it was shown that the thermodynamic properties of the intermediate regime differ from those of sparsely and densely diluted mean-field models.
These probability distribution-dependent differences are thought to be related to the validity of the central limit theorem~\cite{MP}.

In the present study, we rigorously prove that the free energy of nonsparsely diluted mean-field models ($0<b\le1$) with appropriate parameterization is exactly equal to that of the corresponding mean-field models for both ferromagnetic and spin-glass models composed of an arbitrary discrete spin in the thermodynamic limit when the existence probability of each edge follows a Bernoulli distribution.
Thus, it is sufficient to analyze the corresponding mean-field models to investigate the thermodynamic properties of nonsparsely diluted mean-field models.
Our result is a broad generalization of a previous study by Bovier and Gayrard for the densely diluted Curie--Weiss model~\cite{BG}.
The present proof is based on the free energy equivalence between sparsely diluted mean-field models in the infinite connectivity limit and the corresponding mean-field models using the interpolation method~\cite{GT,SG}.

The remainder of this paper is organized as follows.
In Sec. II, we define the model and present the main results (Theorem 1).
Section III presents the proof of Theorem 1. 
Finally, a discussion is presented in Sec. IV.

\section{Models and result }
We define the nonsparsely diluted mean-field ferromagnetic model as
\be
H_{\text{dMF, F}}&=&-\frac{p!}{2\alpha  N^{b(p-1)} } \sum_{1\le i_1<\cdots<i_p \le N} K_{i_1\cdots i_p}  S_{i_1}\cdots S_{i_p} , \label{dilute-mean-ferro}
\ee
where $0<\alpha\le1$, $0<b\le 1$, $p$ is any positive integer, $N$ is the system size, the spin $S_{i}$ takes any bounded discrete value with $|S_i|\le C<\infty$, $p!/(2\alpha  N^{b(p-1)})$ represents the strength of the $p$-body interaction $S_{i_1}\cdots S_{i_p}$,
and $K_{i_1\cdots i_p}$ represent the dilution of interactions and are independent and identically distributed (i.i.d.) random variables following a Bernoulli distribution, with 
\be
\Pr[K_{i_1\cdots i_p}=1]&=&\alpha N^{(b-1)(p-1)},
\\
\Pr[K_{i_1\cdots i_p}=0]&=&1-\alpha N^{(b-1)(p-1)},
\\
\mathbb{E}[K_{i_1\cdots i_p}]&=&\alpha N^{(b-1)(p-1)}, \label{exp-dilute}
\ee
where $\mathbb{E}\left[\cdots\right]$ denotes the expectation with respect to all the random variables.
Since the expectation value of $K_{i_1\cdots i_p}$ is $\mathcal{O} ( N^{(b-1)(p-1)})$, the interaction strength is $\mathcal{O} ( N^{-b(p-1)})$ to satisfy the extensive property in the thermodynamic limit.  
The parameter $\alpha$ in the denominator of the interaction strength in Eq. (\ref{dilute-mean-ferro}) is introduced to simplify the corresponding mean-field ferromagnetic model defined below, which is referred to as appropriate parameterization in the present study.
Note that $b=1$ and $0<b<1$ correspond to the densely diluted mean-field model and intermediately diluted mean-field model, respectively (the case $b=0$ corresponds to sparsely diluted mean-field models but is not treated in the present study).

Similarly, the nonsparsely diluted mean-field spin-glass model is defined as follows:
\be
H_{\text{dMF, SG}}&=&-\sqrt{\frac{p!}{2 \alpha N^{b(p-1)} } } \sum_{1\le i_1<\cdots<i_p \le N} K_{i_1\cdots i_p} J_{d,i_1\cdots i_p} S_{i_1}\cdots S_{i_p} ,
\ee
where we consider the following two probability distribution cases of $J_{d,i_1\cdots i_p}$: (i) a Gaussian distribution $\mathcal{N}\left(J_0\sqrt{p!/ (2\alpha  N^{b(p-1)})}, {J^2}\right)$ and (ii) any bounded discrete probability distribution with
\be
\mathbb{E}[J_{d,i_1\cdots i_p}]&=&J_0\sqrt{\frac{p! }{2 \alpha N^{b(p-1)} }} +\mathcal{O}(\alpha^{-1} N^{-b(p-1)}), \label{dis-mean}
\\
\mathbb{E}[J_{d,i_1\cdots i_p}^2]&=&{J^2}  +\mathcal{O}(\alpha^{-1/2} N^{-b(p-1)/2}) , \label{dis-vari}
\\
\mathbb{E}[J_{d,i_1\cdots i_p}^n]&<&\infty \quad (n\ge3), \label{dis-bound}
\ee
where $J>0$ and $J_0\ge0$.
The simplest example of (ii) is the case of a binary distribution with
\be
 \Pr[J_{d,i_1\cdots i_p}=1]&=&\frac{ {J^2}+J_0\sqrt{\frac{p! }{2 \alpha N^{b(p-1)} }} }{2},
 \\
 \Pr[J_{d,i_1\cdots i_p}=-1]&=&\frac{ {J^2}-J_0\sqrt{\frac{p! }{2 \alpha N^{b(p-1)} }} }{2}.
 \ee
In this case, Eqs. (\ref{dis-mean}) and (\ref{dis-vari}) hold without correction term.

The partition functions of the nonsparsely diluted mean-field models are given by
\be
Z_{\text{dMF, F}}&=&\Tr(e^{-\beta H_{\text{dMF, F}}}) ,
\\
Z_{\text{dMF, SG}}&=&\Tr(e^{-\beta H_{\text{dMF, SG}}}) ,
\ee
where $\Tr(\cdots)$ denotes the summation with respect to all spin variables, and $\beta$ is the inverse temperature.
The free energies of the nonsparsely diluted mean-field models are given by
\be
\alpha_{\text{dMF, F}}&=& -\lim_{N\to\infty}\frac{1}{N\beta} \log Z_{\text{dMF, F}},
\\
\alpha_{\text{dMF, SG}}&=& -\lim_{N\to\infty}\frac{1}{N\beta} \log Z_{\text{dMF, SG}}.
\ee 
The quenched free energies of the nonsparsely diluted mean-field models are defined as
\be
f_{\text{dMF, F}}&=& -\lim_{N\to\infty}\frac{1}{N\beta} \mathbb{E}[\log Z_{\text{dMF, F}}],
\\
f_{\text{dMF, SG}}&=& -\lim_{N\to\infty}\frac{1}{N\beta} \mathbb{E}[\log Z_{\text{dMF, SG}}].
\ee
When the probability distribution of $J_{d,i_1\cdots i_p}$ is symmetric, it is possible to prove the existence of the thermodynamic limit of the quenched free energies of mean-field models in some cases~\cite{GT2,FL,BGT}. 
However, no such proof is known in the case of the present study.
Therefore, in the following, we assume the existence of the thermodynamic limit of the quenched free energies of the nonsparsely diluted mean-field models.
Then, by employing the same approach described in the literature~\cite{PS,Guerra}, it is possible to prove that the free energies of the nonsparsely diluted mean-field models have the self-averaging property and converge almost surely to the quenched free energies in the thermodynamic limit.
Thus, it is sufficient to investigate the quenched free energies of the nonsparsely diluted mean-field models.

The Hamiltonians of the corresponding mean-field ferromagnetic and spin-glass models are defined as
\be
H_{\text{MF, F}}&=&-\frac{p!}{2  N^{p-1} } \sum_{1\le i_1<\cdots<i_p \le N}  S_{i_1}\cdots S_{i_p} ,
\\
H_{\text{MF, SG}}&=&-\sqrt{\frac{p!}{2  N^{p-1} }} \sum_{1\le i_1<\cdots<i_p \le N} J_{i_1\cdots i_p} S_{i_1}\cdots S_{i_p} ,
\ee
where $J_{i_1\cdots i_p}$ are i.i.d. random variables following a Gaussian distribution, $\mathcal{N}\left(J_0\sqrt{p!/(2  N^{p-1})}, {J^2}\right)$.
The quenched free energies of the corresponding mean-field models are defined as follows:
\be
f_{\text{MF, F}}&=& -\lim_{N\to\infty}\frac{1}{N\beta} \mathbb{E}[\log Z_{\text{MF, F}}],
\\
f_{\text{MF, SG}}&=& -\lim_{N\to\infty}\frac{1}{N\beta} \mathbb{E}[\log Z_{\text{MF, SG}}].
\ee
Our results are as follows:
\begin{theorem}
Let $\alpha \in (0,1]$ and $b\in (0,1]$ be functions of $N$ such that $\alpha N^{b(p-1)}\to \infty$ is $N\to\infty$.
Then,
\be
f_{\text{MF, F}}&=&f_{\text{dMF, F}},
\\
f_{\text{MF, SG}}&=&f_{\text{dMF, SG}}.
\ee
\end{theorem}

\begin{remark}
The free energy of the densely diluted ferromagnetic Ising model ($b=1$ and $p=2$) coincides exactly with that of the Curie--Weiss model~\cite{BG}.
Theorem 1 extends this result to any bounded discrete spin, $p$-body interaction, intermediately diluted regime ($0<b<1$), and spin-glass model.
\end{remark}

\section{Proof of Theorem 1}
We provide a proof only for the spin-glass model (a similar proof also applies to the ferromagnetic model). 
Our proof is based on the free energy equivalence between sparsely diluted mean-field models ($b=0$) in the infinite connectivity limit and the corresponding mean-field models using the interpolation method~\cite{GT,SG}.
This method introduces a parameter interpolating two models and examines the response of the free energy to that parameter.

For the interpolating parameter $t$ $(0\le t\le1)$, the interpolating Hamiltonian is defined as
\be
H_N(t)&=&- \sqrt{\frac{p!}{2 \alpha N^{b(p-1)} } } \sum_{1\le i_1<\cdots<i_p \le N} K_{i_1\cdots i_p}(t) J_{d,i_1\cdots i_p} S_{i_1}\cdots S_{i_p} -\sqrt{\frac{p!}{2  N^{p-1} }} \sum_{1\le i_1<\cdots<i_p \le N} J_{i_1\cdots i_p}(t) S_{i_1}\cdots S_{i_p}, \label{inter-Hamil}
\no\\
\ee
where $K_{i_1\cdots i_p}(t)$ follows a Bernoulli distribution with $\mathbb{E}[K_{i_1\cdots i_p}(t)]=t\alpha N^{(b-1)(p-1)}$, and $J_{i_1\cdots i_p}(t)$ follows a Gaussian distribution $\mathcal{N}\left((1-t)J_0\sqrt{p!/(2\alpha N^{p-1}) },(1-t){J^2}\right)$.
It is worth noting that $J_{d,i_1\cdots i_p}$ and $J_{i_1\cdots i_p}(t)$ are sampled from different distributions.
The interpolating pressure function is given by
\be
A_N(t)
&=&\frac{1}{N}\mathbb{E}\left[\log \Tr \left( e^{-\beta H_N(t)}  \right) \right] .\label{inter-pressure}
\ee
Note that 
\be
\lim_{N\to\infty}A_N(0)&=&-\beta f_{\text{MF, SG}},
\\
\lim_{N\to\infty}A_N(1)&=&-\beta f_{\text{dMF, SG}}.
\ee
The following relationship is useful for any function $f(x)$
\be
\frac{\mathrm{d}}{\mathrm{d}t} \mathbb{E}[f(K_{i_1\cdots i_p}(t))]&=&\alpha N^{(b-1)(p-1)} \left(f(1) - f(0) \right). \label{use-ident}
\ee
Using Eq. (\ref{use-ident}) and integration by parts, we obtain
\be
\frac{\mathrm{d} A_N(t)}{\mathrm{d}t}
&=&\frac{1 }{N} \mathbb{E}\left[ \alpha N^{(b-1)(p-1)}\sum_{i_1<\cdots<i_p} \log\left\langle e^{\beta \sqrt{\frac{p!}{2 \alpha N^{b(p-1)} }} J_{d,i_1 \cdots i_p} S_{i_1}\cdots S_{i_p}} \right\rangle_{t, K_{i_1\cdots i_p}(t)=0}   \right]
\no\\
&&-\frac{1 }{N} \mathbb{E}\left[ {\frac{\beta p! J_0}{2  N^{p-1} }}  \sum_{i_1<\cdots<i_p} \langle S_{i_1}\cdots S_{i_p} \rangle_t  +  \frac{\beta^2 p!J^2}{4N^{p-1}}\sum_{i_1<\cdots<i_p}\left(\langle S_{i_1}^2\cdots S_{i_p}^2 \rangle_t-\langle S_{i_1}\cdots S_{i_p} \rangle_t^2\right) \right], \no\\\label{interpolation-deri}
\ee
where $\langle\cdots \rangle_{t}$ is the thermal average with respect to the interpolating Hamiltonian (\ref{inter-Hamil})
\be
\langle\cdots \rangle_{t}&=&\frac{\Tr(\cdots e^{-\beta H_N(t)} )}{\Tr(e^{-\beta H_N(t)} ) },
\ee
and $\langle\cdots \rangle_{t, K_{i_1\cdots i_p}(t)=0}$ is the thermal average with respect to the interpolating Hamiltonian (\ref{inter-Hamil}), except for the interaction $ \sqrt{p! /(2\alpha N^{b(p-1)} )}K_{i_1\cdots i_p}(t)J_{d,i_1 \cdots i_p} S_{i_1}\cdots S_{i_p}$
\be
\langle\cdots \rangle_{t, K_{i_1\cdots i_p}(t)=0}&=&\frac{\Tr(\cdots e^{-\beta H_N(t)  -\beta \sqrt{\frac{p!}{2 \alpha N^{b(p-1)} }} K_{i_1\cdots i_p}(t) J_{d,i_1 \cdots i_p} S_{i_1}\cdots S_{i_p}} )}{\Tr(e^{-\beta H_N(t)  -\beta \sqrt{\frac{p!}{2 \alpha N^{b(p-1)} }} K_{i_1\cdots i_p}(t) J_{d,i_1 \cdots i_p} S_{i_1}\cdots S_{i_p}} ) } .
\ee

\subsection{Case where $J_{d,i_1\cdots i_p}$ follows any bounded discrete probability distribution}
As the values of $J_{d,i_1\cdots i_p}$ and $S_{i_j}$ are bounded, the following inequality holds by choosing $N$ to be sufficiently large:
\be
|e^{\beta \sqrt{\frac{p!}{2 \alpha N^{b(p-1)} }} J_{d,i_1 \cdots i_p}| S_{i_1}\cdots S_{i_p}| }-1|
<1. \label{taylor-condi}
\ee
We can then expand the logarithmic function as
\be
&&\log\left\langle e^{\beta \sqrt{\frac{p!}{2 \alpha N^{b(p-1)} }} J_{d,i_1 \cdots i_p} S_{i_1}\cdots S_{i_p} } \right\rangle_{t, K_{i_1\cdots i_p}(t)=0} \label{log-func}
\\
&=&\log\left( 1+  \sum_{n=1}^\infty  \frac{1}{n!}\left(\beta \sqrt{\frac{p!}{2 \alpha N^{b(p-1)} }} J_{d,i_1 \cdots i_p}\right)^n  \langle S_{i_1}^n \cdots S_{i_p}^n \rangle_{t, K_{i_1\cdots i_p}(t)=0} \right)
\no\\
&=&\sum_{l=1}^\infty \frac{ (-1)^{l-1}}{l} \left(\sum_{n=1}^\infty  \frac{1}{n!}\left(\beta \sqrt{\frac{p!}{2 \alpha N^{b(p-1)} }} J_{d,i_1 \cdots i_p}\right)^n  \langle S_{i_1}^n\cdots S_{i_p}^n \rangle_{t, K_{i_1\cdots i_p}(t)=0} \right)^{l}. \label{log-expand}
\ee
From Eqs. (\ref{dis-mean}), (\ref{dis-vari}), and (\ref{dis-bound}), the leading-order term of Eq. (\ref{log-expand}) is as follows: $l=n=1$
\be
\mathbb{E}\left[ \beta \sqrt{\frac{p!}{2 \alpha N^{b(p-1)} }} J_{d,i_1 \cdots i_p}  \langle S_{i_1}\cdots S_{i_p} \rangle_{t, K_{i_1\cdots i_p}(t)=0} \right]
&=&  \mathbb{E}\left[ \beta {\frac{p!}{2 \alpha N^{b(p-1)} }}J_0   \langle S_{i_1}\cdots S_{i_p} \rangle_{t, K_{i_1\cdots i_p}(t)=0} \right],
\ee
$l=1,n=2$
\be
\mathbb{E}\left[ \frac{1}{2}\left(\beta \sqrt{\frac{p!}{2 \alpha N^{b(p-1)} }} J_{d,i_1 \cdots i_p} \right)^2  \langle S_{i_1}^2 \cdots S_{i_p}^2 \rangle_{t, K_{i_1\cdots i_p}(t)=0} \right]
&=&\mathbb{E}\left[\frac{\beta^2 p!  J^2}{4\alpha N^{b(p-1)}}  \langle S_{i_1}^2 \cdots S_{i_p}^2 \rangle_{t, K_{i_1\cdots i_p}(t)=0} \right],
\ee
and $l=2,n=1$
\be
\mathbb{E}\left[\frac{ -1}{2} \left( \beta \sqrt{\frac{p!}{2 \alpha N^{b(p-1)} }} J_{d,i_1 \cdots i_p} \langle S_{i_1}\cdots S_{i_p} \rangle_{t, K_{i_1\cdots i_p}(t)=0} \right)^{2}\right]
&=&\mathbb{E}\left[-  \frac{\beta^2 p!  J^2}{{4\alpha N^{b(p-1)}}}  \langle S_{i_1}\cdots S_{i_p} \rangle_{t, K_{i_1\cdots i_p}(t)=0}^2 \right].
\ee
Thus, we can rewrite Eq. (\ref{interpolation-deri}) as follows:
\be
&&\frac{\mathrm{d}A_N(t)}{\mathrm{d}t}
\no\\
&=&\frac{1 }{N} \mathbb{E}\left[ {\frac{\beta p!J_0}{2  N^{p-1} }}  \sum_{i_1<\cdots<i_p} (\langle S_{i_1}\cdots S_{i_p} \rangle_{t, K_{i_1\cdots i_p}(t)=0} -\langle S_{i_1}\cdots S_{i_p} \rangle_t)  \right]
\no\\
&&+\frac{1 }{N} \mathbb{E}\left[ \frac{\beta^2 p!J^2}{4N^{p-1}} \sum_{i_1<\cdots<i_p}\left( \langle S_{i_1}^2\cdots S_{i_p}^2 \rangle_{t, K_{i_1\cdots i_p}(t)=0}-\langle S_{i_1}\cdots S_{i_p} \rangle_{t, K_{i_1\cdots i_p}(t)=0}^2- \langle S_{i_1}^2\cdots S_{i_p}^2 \rangle_t + \langle S_{i_1}\cdots S_{i_p} \rangle_t^2 \right) \right]
\no\\
&&+  \mathcal{O}(\alpha^{-1/2} N^{-b(p-1)/2})   .
\ee
Furthermore, it is easy to verify the following:
\be
\mathbb{E}[\langle\cdots \rangle_t]&=&\mathbb{E}[ \langle\cdots \rangle_{t, K_{i_1\cdots i_p}(t)=0} ]+ \mathcal{O}(\alpha^{-1/2} N^{-b(p-1)/2}),
\ee  
because the taylor expansion implies
\be
\langle\cdots \rangle_t&=&\frac{\Tr(\cdots e^{-\beta H(t)})}{\Tr( e^{-\beta H(t)})}
\no\\
&=&\frac{\langle \cdots e^{\beta\sqrt{\frac{p!}{2 \alpha N^{b(p-1)} } }  K_{i_1\cdots i_p}(t) J_{d,i_1\cdots i_p} S_{i_1}\cdots S_{i_p}  }\rangle_{t, K_{i_1\cdots i_p}(t)=0} }{\langle e^{\beta\sqrt{\frac{p!}{2 \alpha N^{b(p-1)} } }  K_{i_1\cdots i_p}(t) J_{d,i_1\cdots i_p} S_{i_1}\cdots S_{i_p}  }\rangle_{t, K_{i_1\cdots i_p}(t)=0} }
\no\\
&=&\frac{\langle \cdots \rangle_{t, K_{i_1\cdots i_p}(t)=0} +\mathcal{O}(\alpha^{-1/2} N^{-b(p-1)/2}) }{1+\mathcal{O}(\alpha^{-1/2} N^{-b(p-1)/2})}
\no\\
&=& \langle\cdots \rangle_{t, K_{i_1\cdots i_p}(t)=0} + \mathcal{O}(\alpha^{-1/2} N^{-b(p-1)/2}).
\ee
Consequently, we obtain
\be
\frac{\mathrm{d}A_N(t)}{\mathrm{d}t}
&=&\mathcal{O}(\alpha^{-1/2} N^{-b(p-1)/2}) .
\ee
Finally, taking the thermodynamic limit $N\to\infty$ such that $\alpha N^{b(p-1)}\to \infty$, we obtain
\be
\lim_{N\to\infty}A_N(1)&=&\lim_{N\to\infty} \left( A_N(0)+ \int_0^1dt \frac{\mathrm{d}A_N(t)}{\mathrm{d}t}\right)
\no\\
&=&\lim_{N\to\infty} \left( A_N(0)+ \int_0^1dt \mathcal{O}(\alpha^{-1/2} N^{-b(p-1)/2}) \right)
\no\\
&=&\lim_{N\to\infty}  A_N(0).
\ee
This is the proof of Theorem 1.

\subsection{Case  where $J_{d,i_1\cdots i_p}$ follows the Gaussian distribution}
The procedure of the proof is almost the same but we have to pay attention to expanding the logarithmic  function (\ref{log-func}).
Since $J_{d,i_1\cdots i_p}$ is defined on $(-\infty,\infty)$, Eq. (\ref{log-func}) cannot be expanded when $J_{d,i_1\cdots i_p}$ is large.
Indeed, this is not a problem, because the contribution from regions with large $J_{d,i_1\cdots i_p}$ becomes exponentially small in Eq. (\ref{log-func}) as follows (see Appendix A for the derivation)
\be
\mathcal{O}\left(e^{-\left(\log{2}/(\beta C^p) \right)^2 \alpha N^{b(p-1)} /(p!J^2)}\right).  \label{extreme-value}
\ee
Thus, we can neglect the contribution from regions with large $J_{d,i_1\cdots i_p}$ in Eq. (\ref{log-func}).
Then, we can expand the logarithmic function (\ref{log-func}) as in the case of bounded discrete probability distribution, and the calculations are the same.

\section{Discussions}
We rigorously proved that the free energies of the densely diluted mean-field models ($b=1$) and intermediately diluted mean-field models ($0< b<1$) with appropriate parameterization exactly coincide with that of the corresponding mean-field models in both ferromagnetic and spin-glass models composed of any discrete spin $S$ in the thermodynamic limit when the existence probability of each edge follows a Bernoulli distribution.

Note that the value of $\alpha$ is allowed to be close to zero, as long as the condition $\alpha N^{b(p-1)}\to \infty$ is satisfied within the thermodynamic limit.
This explains why the ground-state energy of the densely diluted SK model ($b=1$ and $p=2$) coincides with that of the SK model, and depends neither on the distribution of interactions nor on the dilution concentration, even within the limit $\alpha\to0$~\cite{Boettcher}.

Furthermore, our results determine the thermodynamic properties of intermediately diluted mean-field models ($0< b<1$) at finite temperatures.
This rigorously confirms recent studies~\cite{MP,FM}, where the thermodynamic properties of the intermediate regime coincide with those of the corresponding mean-field models if the existence probability of each edge follows a Bernoulli distribution.
On the other hand, the previous study~\cite{MP} suggests that this equivalence does not hold when the existence probability of each edge follows an exponential distribution. In this case, our method does not work well, because the derivative of the interpolation parameter (\ref{use-ident}) does not take a simple form.

\section*{Data availability statement}
No new data were created or analysed in this study.

\section*{Acknowledgment}
The authors are grateful to Fernando Lucas Metz for useful comments.
This study was supported by JSPS KAKENHI Grant Nos. 24K16973 and 23H01432.
Our study received financial support from the Public\verb|\|Private R\&D Investment Strategic Expansion PrograM (PRISM) and programs for bridging the gap between R\&D and IDeal society (Society 5.0) and Generating Economic and social value (BRIDGE) from the Cabinet Office.

\appendix
\section{Derivation of Eq. (\ref{extreme-value})} 
Since $|S_i|\le C<\infty$, we find
\be
\exp\left(\beta \sqrt{\frac{p!}{2 \alpha N^{b(p-1)} }} J_{d,i_1 \cdots i_p} S_{i_1}\cdots S_{i_p} \right)
&\le& \exp\left(\beta \sqrt{\frac{p!}{2 \alpha N^{b(p-1)} }} |J_{d,i_1 \cdots i_p}| C^p\right) .
\ee
We define $x_N>0$ such that 
\be
\exp\left(\beta \sqrt{\frac{p!}{2 \alpha N^{b(p-1)} }} x_N C^p\right) -1=1,\label{x-difi}
\ee
equivalently,
\be
x_N&=& \frac{\log2}{\beta C^p}\sqrt{\frac{2 \alpha N^{b(p-1)} }{p!} }. 
\ee
We take the expectatin of Eq. (\ref{log-func}) and divide the integral interval of $J_{d,i_1 \cdots i_p}$  into three parts
\be
&& \mathbb{E}\left[  \log\left\langle e^{\beta \sqrt{\frac{p!}{2 \alpha N^{b(p-1)} }} J_{d,i_1 \cdots i_p} S_{i_1}\cdots S_{i_p}} \right\rangle_{t, K_{i_1\cdots i_p}(t)=0}  \right]
\no\\
&=& \mathbb{E}'\left[\left( \int_{-\infty}^{-x_N} +\int_{-x_N}^{x_N} +\int_{x_N}^{\infty}\right)dJ_{d,i_1 \cdots i_p} \frac{e^{-\frac{\left(J_{d,i_1 \cdots i_p}-J_0\sqrt{\frac{p!}{2\alpha  N^{b(p-1)}) }}\right)^2}{2J^2}}}{\sqrt{2\pi}} \log\left\langle e^{\beta \sqrt{\frac{p!}{2 \alpha N^{b(p-1)} }} J_{d,i_1 \cdots i_p} S_{i_1}\cdots S_{i_p}} \right\rangle_{t, K_{i_1\cdots i_p}(t)=0}   \right],
\no\\
 \label{three-inter}
\ee
where $\mathbb{E}'[\cdots]$ denotes the expectation with respect to all the random variables except for $J_{d,i_1 \cdots i_p}$.
From Eq. (\ref{x-difi}), the integrand in Eq. (\ref{three-inter}) can be expanded in the interval $-x_N<J_{d,i_1 \cdots i_p}<x_N$.

Our purpose is to show that the contribution from the two intervals $-\infty<J_{d,i_1 \cdots i_p}\le -x_N$ and  $x_N\le J_{d,i_1 \cdots i_p}<\infty$ in Eq. (\ref{three-inter}) is exponentially small for the system size $N$.
In the interval $x_N\le J_{d,i_1 \cdots i_p}<\infty$, we evaluate
\be
&&\left| \int_{x_N}^{\infty}dJ_{d,i_1 \cdots i_p} \frac{e^{-\frac{\left(J_{d,i_1 \cdots i_p}-J_0\sqrt{\frac{p!}{2\alpha  N^{b(p-1)}) }}\right)^2}{2J^2}}}{\sqrt{2\pi}} \log\left\langle e^{\beta \sqrt{\frac{p!}{2 \alpha N^{b(p-1)} }} J_{d,i_1 \cdots i_p} S_{i_1}\cdots S_{i_p}} \right\rangle_{t, K_{i_1\cdots i_p}(t)=0}   \right|
\no\\
&\le& \int_{x_N}^{\infty}dJ_{d,i_1 \cdots i_p} \frac{e^{-\frac{\left(J_{d,i_1 \cdots i_p}-J_0\sqrt{\frac{p!}{2\alpha  N^{b(p-1)}) }}\right)^2}{2J^2}}}{\sqrt{2\pi}} \log\left\langle e^{\beta \sqrt{\frac{p!}{2 \alpha N^{b(p-1)} }} J_{d,i_1 \cdots i_p} C^p} \right\rangle_{t, K_{i_1\cdots i_p}(t)=0}   
\no\\
&=& \beta \sqrt{\frac{p!}{2 \alpha N^{b(p-1)} }} C^p  \int_{x_N-J_0\sqrt{\frac{p!}{2\alpha  N^{b(p-1)} }}}^{\infty}dJ_{d,i_1 \cdots i_p} \frac{e^{-\frac{\left(J_{d,i_1 \cdots i_p}\right)^2}{2J^2}}}{\sqrt{2\pi}} \left(J_{d,i_1 \cdots i_p} +J_0\sqrt{\frac{p!}{2\alpha  N^{b(p-1)} }} \right)  
\no\\
&=& \beta \sqrt{\frac{p!}{2 \alpha N^{b(p-1)} }} C^p \frac{J^2 e^{-\frac{1}{2J^2}(x_N-J_0\sqrt{\frac{p!}{2\alpha  N^{b(p-1)} }} )^2 }}{\sqrt{2\pi}}
\no\\
&&+\beta \frac{p!}{2 \alpha N^{b(p-1)} }C^pJ_0  \int_{x_N-J_0\sqrt{\frac{p!}{2\alpha  N^{b(p-1)} }} }^{\infty}dJ_{d,i_1 \cdots i_p} \frac{e^{-\frac{\left(J_{d,i_1 \cdots i_p}\right)^2}{2J^2}}}{\sqrt{2\pi}} .\label{+-inberval}
\ee
Furthermore, it is known that the second term in Eq. (\ref{+-inberval}) is bounded by 
\be
\int_{x_N-J_0\sqrt{\frac{p!}{2\alpha  N^{b(p-1)} }} }^{\infty}dJ_{d,i_1 \cdots i_p} \frac{e^{-\frac{\left(J_{d,i_1 \cdots i_p}\right)^2}{2J^2}}}{\sqrt{2\pi}} 
&\le&\frac{J^2}{x_N-J_0\sqrt{\frac{p!}{2\alpha  N^{b(p-1)} }}} \frac{1}{\sqrt{2\pi}} e^{-\frac{1}{2J^2} \left(x_N-J_0\sqrt{\frac{p!}{2\alpha  N^{b(p-1)} }} \right)^2 } , \label{Mills ratio}
\ee
which is called the Mills ratio~\cite{Grimmett}.
From Eqs. (\ref{+-inberval}) and (\ref{Mills ratio}), we obtain
\be
&& \int_{x_N}^{\infty}dJ_{d,i_1 \cdots i_p} \frac{e^{-\frac{\left(J_{d,i_1 \cdots i_p}-J_0\sqrt{\frac{p!}{2\alpha  N^{b(p-1)}) }}\right)^2}{2J^2}}}{\sqrt{2\pi}} \log\left\langle e^{\beta \sqrt{\frac{p!}{2 \alpha N^{b(p-1)} }} J_{d,i_1 \cdots i_p} S_{i_1}\cdots S_{i_p}} \right\rangle_{t, K_{i_1\cdots i_p}(t)=0}   
\no\\
&\le&\beta \sqrt{\frac{p!}{2 \alpha N^{b(p-1)} }} C^p \frac{J^2e^{-\frac{1}{2J^2}(x_N-J_0\sqrt{\frac{p!}{2\alpha  N^{b(p-1)} }} )^2 }}{\sqrt{2\pi}}
\no\\
&&+\beta \frac{p!}{2 \alpha N^{b(p-1)} }C^pJ_0  \frac{J^2}{x_N-J_0\sqrt{\frac{p!}{2\alpha  N^{b(p-1)} }}} \frac{1}{\sqrt{2\pi}} e^{-\frac{1}{2J^2} \left(x_N-J_0\sqrt{\frac{p!}{2\alpha  N^{b(p-1)} }} \right)^2 } 
\no\\
&=&\mathcal{O}(e^{-x_N^2/(2J^2)})
\no\\
&=&\mathcal{O}\left(e^{-\left(\log{2}/(\beta C^p) \right)^2 \alpha N^{b(p-1)} /(p!J^2)}\right) .
\ee
This implies that the contribution from the interval $x_N\le J_{d,i_1 \cdots i_p}<\infty$ in Eq. (\ref{three-inter}) is exponentially small for the system size $N$.
A similar calculation is applicable to the interval $-\infty<J_{d,i_1 \cdots i_p}\le -x_N$.
Thus, Eq. (\ref{three-inter}) is rewritten as
\be
&& \mathbb{E}\left[  \log\left\langle e^{\beta \sqrt{\frac{p!}{2 \alpha N^{b(p-1)} }} J_{d,i_1 \cdots i_p} S_{i_1}\cdots S_{i_p}} \right\rangle_{t, K_{i_1\cdots i_p}(t)=0}  \right]
\no\\
&=& \mathbb{E}'\left[ \int_{-x_N}^{x_N} dJ_{d,i_1 \cdots i_p} \frac{e^{-\frac{\left(J_{d,i_1 \cdots i_p}-J_0\sqrt{\frac{p!}{2\alpha  N^{b(p-1)}) }}\right)^2}{2J^2}}}{\sqrt{2\pi}} \log\left\langle e^{\beta \sqrt{\frac{p!}{2 \alpha N^{b(p-1)} }} J_{d,i_1 \cdots i_p} S_{i_1}\cdots S_{i_p}} \right\rangle_{t, K_{i_1\cdots i_p}(t)=0}   \right]
\no\\
&&+\mathcal{O}\left(e^{-\left(\log{2}/(\beta C^p) \right)^2 \alpha N^{b(p-1)} /(p!J^2)}\right) ,
\ee
and the logarithmic function can be expanded in the interval $-x_N<J_{d,i_1 \cdots i_p}<x_N$.


\end{document}